\documentclass[11pt, draftcls, onecolumn]{IEEEtran}
\usepackage[english]{babel}
\usepackage{amsmath,amsthm}
\usepackage{amsfonts}
\usepackage{graphicx}
\usepackage{verbatim}
\usepackage{amssymb}
\usepackage{epstopdf}
\usepackage{hyperref}
\usepackage{bbm}
\usepackage{cite,bbm,graphicx,amsmath,amssymb,mathrsfs,epsf} \usepackage{multirow}
\usepackage{subfigure,cite,graphicx,epsfig,amsmath,amssymb,mathrsfs,epsf,graphics,color,enumerate}

\newtheorem{thm}{Theorem}%[section]

\newtheorem{rem}{Remark}
\def \E{\operatorname{E}}

\begin{document}
\title{Secure Degrees of Freedom of the Gaussian Diamond-Wiretap Channel}%
\author{\IEEEauthorblockN{Si-Hyeon Lee, Wanyao Zhao, and Ashish Khisti}\thanks{S.-H. Lee and A. Khisti are with the Department of Electrical and Computer Engineering,
University of Toronto, Toronto, Canada (e-mail: sihyeon.lee@utoronto.ca; akhisti@comm.utoronto.ca). W. Zhao was at University of Toronto when this work was done.  This work was supported by QNRF, a member
 of Qatar Foundation, under NPRP project 5-401-2-161. }}
\maketitle

\begin{abstract}
In this paper, we consider the Gaussian diamond-wiretap channel that consists of an orthogonal broadcast channel from a source to two relays and a Gaussian fast-fading multiple access-wiretap channel from the two relays to a legitimate destination and an eavesdropper. For the multiple access part, we consider both the case with full channel state information (CSI) and the case with no eavesdropper's CSI, at the relays and the legitimate destination. For both the cases, we establish the exact secure degrees of freedom and generalize the results for multiple relays. 

For the converse part, we introduce a new technique of capturing the trade-off between the message rate and the amount of individual randomness injected at each relay. In the achievability part, we show (i) how to strike a balance between sending message symbols and common noise symbols from the source to the relays in the broadcast component and (ii) how to combine artificial noise-beamforming and noise-alignment techniques at the relays in the multiple access component. In the case with full CSI, we propose a scheme where the relays simultaneously beamform common noise signals in the null space of the legitimate destination's channel, and align them with the message signals at the eavesdropper. In the case with no eavesdropper's CSI, we present a scheme that efficiently utilizes the broadcast links by incorporating computation between the message and common noise symbols at the source. Finally, most of our achievability and converse techniques can also be adapted to the Gaussian (non-fading) channel model.

\end{abstract}
% ----------------------------------------------------------------
\section{Introduction}
A model of wiretap channel was first studied by Wyner \cite{Wyner:75}, where a source wishes to send its message to a legitimate destination while keeping it secret from an eavesdropper. Wyner established the secrecy capacity for the degraded case where the eavesdropper receives a physically degraded version of the channel output at the legitimate destination. Csisz{\'{a}}r and K{\"{o}}rner generalized his work to general, not necessarily degraded, discrete memoryless wiretap channel \cite{CsiszarKorner:78}. 
This line of work has been subsequently extended to various multi-user scenarios, see e.g.,  \cite{KoyluogluKoksalElGamal:12,relay-3,LiangPoor:08, relay-4, relay-2,   TekinYener:08,TekinYener:08general,GoelNegi:08, perron_thesis,relay-5, relay-6,YassaeeArefGohari:14,relay-7}, however, the characterization of the secrecy capacity remains a challenging open problem in general.  In fact, even for the seemingly simple case of the Gaussian multiple access-wiretap channel, the secrecy capacity is only known for the degraded case~\cite{TekinYener:08}.

Recently, as an alternative but insightful measure, the secure degrees of freedom (d.o.f.) has been actively studied \cite{HeYener:14, XieUlukus:14,XieUlukus:15,MukherjeeXieUlukus:arxiv15} for various multi-user wiretap networks. For the Gaussian multiple access-wiretap channel, the sum secure d.o.f. was shown to be $\frac{2}{3}$ for almost all channel gains \cite{XieUlukus:14}. For achievability, a cooperative jamming scheme was proposed that incorporates real interference alignment \cite{MotahariOveisGharanMaddahAliKhandani:14} at the legitimate destination and the eavesdropper.  In many practical scenarios, however, it is hard for the source and the legitimate destination to know the eavesdropper's channel state information (CSI). In \cite{MukherjeeXieUlukus:arxiv15}, the secure d.o.f. with no eavesdropper's CSI was characterized for some interesting one-hop wiretap channels. For the Gaussian multiple access-wiretap channel, the sum secure d.o.f. was shown  in \cite{MukherjeeXieUlukus:arxiv15} to reduce to $\frac{1}{2}$ with no eavesdropper's CSIT, which is achieved by a blind cooperative jamming scheme. We note that the prior work has focused on one-hop wiretap networks, and to the best of our knowledge, there has been no prior work on the secure d.o.f. for multi-hop wiretap networks.

\begin{figure}
  \centering
    \includegraphics[width=95mm]{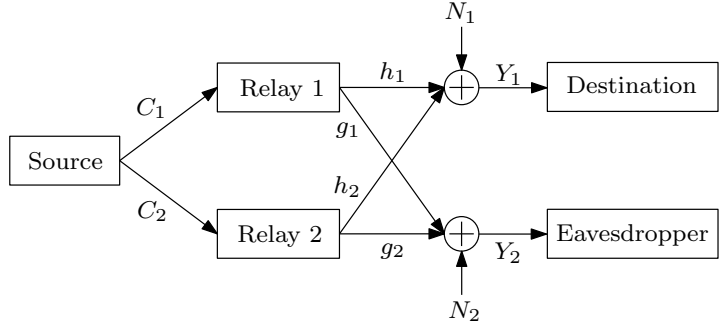}\\
  \caption{The Gaussian diamond-wiretap channel. }\label{fig:model}
\end{figure}

In this paper, we consider the Gaussian diamond-wiretap channel illustrated in Fig. \ref{fig:model} that consists of an orthogonal broadcast channel from a source to two relays and a Gaussian multiple access-wiretap channel from the two relays to a legitimate destination and an eavesdropper. 
We consider both the case where the relays and the legitimate destination know the legitimate CSI and the eavesdropper's CSI and the case where they know only the legitimate CSI, which we call the case with full CSI and the case with no eavesdropper's CSI, respectively.\footnote{We assume that the source does not know any of the legitimate CSI and the eavesdropper's CSI and the eavesdropper knows both the CSI's.} The proposed setting is a two-hop communication network and involves several new elements not present in the single-hop networks studied previously. 
Our model introduces a new possibility of utilizing common message and/or common noise for the Gaussian multiple access-wiretap channel. This brings in an interesting tension in the use of the broadcast links regarding whether we  send independent messages,  common message, common noise, or a function of those across the broadcast part. 
At one extreme, when the capacities of the orthogonal links in the broadcast part are sufficiently small, the optimal strategy turns out to send independent partial message to each relay and incorporate jamming schemes  \cite{XieUlukus:14,MukherjeeXieUlukus:arxiv15} for the multiple access part.
At the other extreme, when the broadcast part has sufficiently high capacity to transmit common message or common noise symbols to the relays, without incurring bottleneck, it follows that the secure d.o.f. equals $1$ using the results \cite{KhistiWornell:10,LiuLiuPoorShamai:10,MukherjeeTandonUlukus:15} for the multiple-input single-output (MISO) wiretap channel. 
When the link capacities of the broadcast part are moderate, however, the optimal scheme is not immediate. 
Furthermore, due to the possibility of sending common information across the broadcast part, we cannot assume in proving converse that the channel inputs and outputs at the relays are independent, whereas channel inputs at transmitters are inherently independent in most one-hop wiretap networks. 

On the other hand, with no secrecy constraint, our model falls back to the  diamond channel introduced by Schein \cite{schein_thesis}, whose capacity is not known in general. For a range of moderate link capacities at the broadcast part, \cite{TraskovKramer:07}, \cite{KangLiu:11} characterized the capacity of the diamond channel, which is strictly tighter than the cutset bound. For achievability, a coding scheme incorporating multicoding at the source was proposed in \cite{TraskovKramer:07}, \cite{KangLiu:11}. For converse, \cite{KangLiu:11} used a technique from \cite{Ozarow:80} to take into account the correlation between the two relay signals. In the presence of a secrecy constraint, such converse proof techniques need to be adopted carefully by taking into account the stochastic encoding functions introduced to confuse the eavesdropper. Those works \cite{TraskovKramer:07}, \cite{KangLiu:11} were generalized in \cite{LeeKhisti:15} for the degraded Gaussian diamond-wiretap channel, in which the secrecy capacity was characterized for several ranges of channel parameters. For non-degraded case, however, the coding scheme used in \cite{LeeKhisti:15} achieves zero secure d.o.f. and structured codes such as interference alignment and beamforming schemes need to be involved to achieve a positive secure d.o.f.

For the  Gaussian diamond-wiretap channel in Fig. \ref{fig:model}, we establish the exact secure d.o.f. in terms of the link d.o.f.'s at the broadcast part, both for the case with full CSI and for the case with no eavesdropper's CSI. We assume a fast fading scenario where channel fading coefficients are i.i.d. across the time, but our converse result for the former case and achievability results for both the cases continue to hold when the channel gains are fixed. For the converse part, we combine the proof techniques in \cite{XieUlukus:14,MukherjeeXieUlukus:arxiv15,DavoodiJafar:14} with a new technique capturing the trade-off between the message rate and the amount of individual randomness injected at each relay. Our achievability part is based on five key constituent schemes. In particular, we propose two new schemes that utilize common noise, in a way that the common noise signals are beam-formed in the null space of the legitimate destination's channel. 
One of these two schemes is for the case with full CSI and is called a simultaneous alignment and beamforming (S-AB) scheme, which incorporates alignment of the message and the common noise signals at the eavesdropper. 
The proposed S-AB scheme also extends easily to the case with more than two relays and yields the best achievable secure d.o.f. 
The other scheme is for the case with no eavesdropper's CSI and is called a computation for jamming (CoJ) scheme, which efficiently utilizes the broadcast links by incorporating computation between the message and the common noise symbols at the source. The remaining three schemes are straightforward extensions of the previously known schemes, i.e., the cooperative jamming scheme \cite{XieUlukus:14} and blind cooperative jamming scheme  \cite{MukherjeeXieUlukus:arxiv15} for the Gaussian multiple access-wiretap channel and the message-beamforming scheme \cite{KhistiWornell:10,LiuLiuPoorShamai:10} for the Gaussian MISO wiretap channel.

As a natural extension, we also consider a generalized Gaussian diamond-wiretap channel with more than two relays. For the brevity of the results, we consider the symmetric case where the link d.o.f.'s of the broadcast part are the same. By generalizing the proof techniques used in the two-relay case, we establish the exact secure d.o.f. for the case with no eavesdropper's CSI and present upper and lower bounds on the secure d.o.f. for the case with full CSI. 

The remaining of this paper is organized as follows. In Section \ref{sec:model}, we formally present the model of the Gaussian diamond-wiretap channel. Our main results on the secure d.o.f. are given in Section \ref{sec:main}. In Sections \ref{sec:converse} and \ref{sec:achievability}, we prove the converse and the achievability parts, respectively. We extend the results for the case with multiple relays in Section \ref{sec:multi}. We conclude this paper in Section \ref{sec:conclusion}.

\section{System Model}\label{sec:model}
Consider the Gaussian diamond-wiretap channel illustrated in Fig.  \ref{fig:model} that consists of a broadcast channel from a source to two relays and a Gaussian multiple access-wiretap channel from the two relays to a legitimate destination and an eavesdropper. For the broadcast part, the source is connected to the two relays through orthogonal links of capacities $C_1$ and $C_2$. For the multiple access part, the channel outputs $Y_1(t)$ and $Y_2(t)$ at time $t$ at the legitimate destination and the eavesdropper, respectively, are given as 
\begin{align}
Y_1(t) &= h_1(t) X_1(t) + h_2(t) X_2(t) + N_1(t) \label{eqn:model_y1}\\
Y_2(t) &= g_1(t) X_1(t) + g_2(t) X_2(t) + N_2(t),\label{eqn:model_y2}
\end{align}
where $X_1(t)$ and $X_2(t)$ are the channel inputs from relays 1 and 2, respectively, $h_k(t)$ and $g_k(t)$ for $k=1,2$ are the channel fading coefficients to the legitimate destination and the eavesdropper, respectively, and $N_1(t)$ and $N_2(t)$ are independent Gaussian noise with zero mean and unit variance at the legitimate destination and the eavesdropper, respectively, at time $t$. The transmit power constraint at relay $k=1,2$ is given as $\frac{1}{n}\sum_{t=1}^nX_{k}^2(t)\leq P$, where $n$ denotes the number of channel uses. 

We assume a fast fading scenario where $h_1(t)$, $h_2(t)$, $g_1(t)$, and $g_2(t)$ are drawn in an i.i.d. fashion over time according to an arbitrary real-valued joint density function $f(h_1, h_2, g_1, g_2)$, whose all joint and conditional density functions are bounded and whose support set does not include zero and infinity, i.e., there exists a positive finite $L$ such that 
\begin{align}
\frac{1}{L}\leq |h_k(t)|, |g_k(t)| \leq L. \label{eqn:ch_assump}
\end{align}
We note that \eqref{eqn:ch_assump} is a mild technical condition because by choosing $L$ large enough, the omitted support set can be reduced to a negligible probability that has a vanishing impact on the degrees of freedom. For notational convenience, let  $\mathbf{h}^t=[h_1(1)~ h_2(1)~ \cdots~ h_1(t)$ ~$h_2(t)]$ and  $\mathbf{g}^t=[g_1(1)~ g_2(1)~ \cdots ~g_1(t) ~g_2(t)]$ denote the legitimate channel state information (CSI) and the eavesdropper's CSI up to time $t$, respectively.

We assume that the source does not know any of the legitimate CSI and the eavesdropper's CSI and the eavesdropper knows both the CSI's.  Two cases are considered regarding the availability of CSI at the relays and the legitimate destination. First, we consider a case where both the legitimate CSI and the eavesdropper's CSI  are available at the two relays and the legitimate destination, which we call the case with full CSI. We also consider another case where only the legitimate CSI is available at the two relays and the legitimate destination, which we call a case with no eavesdropper's CSI.  We note that our achievability and converse techniques for the case with full CSI and our achievability technique for the case with no eavesdropper's CSI can be adapted for the scenario with fixed channel gains over time and for the scenario with complex channel fading coefficients, as remarked at the end of Section \ref{sec:main}.

A $(2^{nR}, n)$ secrecy code consists of a message $W\sim \mbox{Unif}[1:2^{nR}]$,\footnote{$[i:j]$ for two integers $i$ and $j$ denotes the set $\{i,i+1,\cdots, j\}$ and $\mbox{Unif}[S]$ for a set $S$ denotes the uniform distribution over $S$. When $S=[i:j]$, we use $\mbox{Unif}[i:j]$ instead of $\mbox{Unif}[[i:j]]$.} a stochastic encoder at the source that (randomly) maps $W\in [1:2^{nR}]$ to $(J_1, J_2)\in [1:2^{nC_1}]\times [1:2^{nC_2}]$, a stochastic encoder at time $t=1,\ldots,n$ at relay $k=1,2$ that (randomly) maps $(J_k, \mathbf{h}^t, \mathbf{g}^t)$ and $(J_k, \mathbf{h}^t)$ for the case with full CSI and  for the case with no eavesdropper's CSI, respectively, to $X_k(t) \in \mathcal{X}_k$, and a decoding function at the legitimate destination that maps $(Y_1^n,\mathbf{h}^n, \mathbf{g}^n)$ and $(Y_1^n, \mathbf{h}^n)$ for the case with full CSI and for the case with no eavesdropper's CSI, respectively, to $\hat{W} \in [1:2^{nR}]$. The probability of error is given as  $P_e^{(n)}=P(\hat{W}\neq W)$. A secrecy rate of $R$ is said to be achievable if there exists a sequence of $(2^{nR},n)$ codes such that 
$\lim_{n\rightarrow \infty}P_e^{(n)}=0$ and $\lim_{n\rightarrow \infty}\frac{1}{n}I(W;Y_2^n|\mathbf{h}^n, \mathbf{g}^n)=0$.\footnote{Note that there is no secrecy constraint at the relays.}  The secrecy capacity is the supremum of all achievable secrecy rates.

In this paper, we are interested in asymptotic behavior of the secrecy capacity when $P$ tends to infinity. We say a d.o.f. tuple $(\alpha_1, \alpha_2, d_s)$ is achievable if a rate $R$ with $d_s=\lim_{P\rightarrow \infty} \frac{R}{\frac{1}{2}\log P}$ is achievable when $C_1$ and $C_2$ satisfy 
\begin{align*}
\alpha_1=\lim_{P\rightarrow \infty} \frac{C_1}{\frac{1}{2}\log P},~\alpha_2=\lim_{P\rightarrow \infty} \frac{C_2}{\frac{1}{2}\log P}. 
\end{align*}
A secure d.o.f. $d_s(\alpha_1, \alpha_2)$ is the maximum $d_s$ such that $(\alpha_1, \alpha_2, d_s)$ is achievable.  For brevity, $d_s$ denotes $d_s(\alpha_1, \alpha_2)$ according to the context. Without loss of generality, let us assume $C_1\geq C_2$, which implies $\alpha_1\geq \alpha_2$.

%%%%%%%%%%%%%%%%%%%%%%%%%%%%%%%%%%%%%%%%%%%%%%%%%%%%%%%%%%%%%%%%%%%%%
%%%%%%%%%%%%%%%%%%%%%%%%%%%%%%%%%%%%%%%%%%%%%%%%%%%%%%%%%%%%%%%%%%%%%

\section{Main Results }\label{sec:main}
In this section, we state our main results of this paper. The following two theorems present the secure d.o.f. of the Gaussian diamond-wiretap channel for the case with full CSI and for the case with no eavesdropper's CSI, respectively, whose proofs are in Section \ref{sec:converse} for the converse parts and in Section \ref{sec:achievability} for the achievability parts. 
\begin{thm}\label{thm:CSI}
The secure d.o.f. of the Gaussian diamond-wiretap channel with full CSI at the relays and the legitimate destination is equal to 
\begin{align}
d_s = \min
\left\{\alpha_1+\alpha_2, \frac{\alpha_2 + 1}{2}, 1\right\}. \label{eqn:CSI}
\end{align}
\end{thm}
\begin{thm}\label{thm:noCSI}
The secure d.o.f. of the Gaussian diamond-wiretap channel with no eavesdropper's CSI at the relays and the legitimate destination is equal to 
\begin{align}
d_s = \min\left\{\alpha_1+\alpha_2, \frac{\alpha_1+\alpha_2+1}{3},\frac{\alpha_2 + 1}{2},1\right\}.\label{eqn:CSI_no}
\end{align}
\end{thm}

We note that the secure d.o.f. of the classical Gaussian wiretap channel is zero. Theorems \ref{thm:CSI} and \ref{thm:noCSI} show that the secure d.o.f. can be greatly improved by deploying relays. First, note that even if $\alpha_2=0$, the secure d.o.f. of $\frac{1}{2}$ is achievable as long as $\alpha_1\geq \frac{1}{2}$, both for the case with full CSI and for the case with no eavesdropper's CSI. This is because relay 2 can act as a helper \cite{MukherjeeXieUlukus:arxiv15} that enables to produce a jamming signal in cooperation with relay 1. We also note that when each of $\alpha_1$ and $\alpha_2$ is higher than or equal to 1, one secure d.o.f. is achievable for both the cases. For the case with full CSI, this is natural from the known results \cite{KhistiWornell:10,LiuLiuPoorShamai:10} for the Gaussian multiple-input single-output (MISO) wiretap channel, where the source has two antennas and each of the legitimate destination and the eavesdropper has one antenna. In this Gaussian MISO wiretap channel, one secure d.o.f. is achievable by beamforming the message signal in the null space of the eavesdropper's channel. Similarly, if the source can send the message with d.o.f. 1 to both the relays for our diamond-wiretap channel, the relays are able to beam-form the message signals in the null space of the eavesdropper's channel. However, with no eavesdropper's CSI, this is not immediate from the known results for the Gaussian MISO wiretap channel. The secure d.o.f. of the Gaussian MISO wiretap channel is still 1 with no eavesdropper's CSI \cite{MukherjeeTandonUlukus:15}, but it is achieved by sending an artificial noise signal in the null space of the legitimate destination in addition to the message signal. To translate this scheme to our diamond-wiretap channel, the source needs to send to the relays common artificial noise as well as (partial) messages, which requires $\alpha_1\geq 1$, $\alpha_2\geq 1$, and $\alpha_1+\alpha_2\geq 3$. To achieve $(\alpha_1, \alpha_2, d_s)=(1,1,1)$, we propose a novel scheme that incorporates \emph{computation} of the message and artificial noise symbols at the source. This scheme involves transmitting a judicious function of the message and noise symbols from the source such that we only require $(\alpha_1, \alpha_2)= (1,1)$, yet accomplish noise-beamforming as discussed above.

For the special case of symmetric link capacities, i.e., $\alpha_1=\alpha_2=\alpha$, the secure d.o.f.'s are given as
\begin{align*}
\min \left\{2\alpha, \frac{\alpha + 1}{2}, 1\right\},  \min \left\{2\alpha, \frac{2\alpha + 1}{3}, 1\right\}
\end{align*}
for the case with full CSI and for the case with no eavesdropper's CSI, respectively. 
Note that the gain in secure d.o.f. with respect to $\alpha$ is double up to $\alpha=\frac{1}{3}$ and $\alpha=\frac{1}{4}$ for the case with full CSI and for the case with no eavesdropper's CSI, respectively. 
In this range, the broadcast part is the bottleneck and hence it is optimal to send independent partial messages to the relays and to incorporate the cooperative jamming \cite{XieUlukus:14} and the blind cooperative jamming \cite{MukherjeeXieUlukus:arxiv15} for the case with full CSI and for the case with no eavesdropper's CSI, respectively. After this threshold value of $\alpha$, the source needs to send some common information (same message or common artificial noise) to achieve a higher secure d.o.f. and this causes the reduction of the gain in secure d.o.f. with respect to $\alpha$. In Section \ref{sec:multi}, we investigate the effect of the absence of the eavesdropper's CSI on the secure d.o.f. for a generalized model with multiple relays. 

\begin{rem}
For the scenario where the channel fading coefficients are fixed during the whole communication, the lower and upper bounds on the secure d.o.f  for the case with full CSI in Theorem \ref{thm:CSI} and the lower bound on the secure d.o.f. for the case with no eavesdropper's CSI in Theorem \ref{thm:noCSI} continue to hold for almost all channel gains. For an upper bound with no eavesdropper's CSI, a key result from \cite{DavoodiJafar:14} used for the upper bound in Theorem \ref{thm:noCSI}, i.e., the entropy of the channel output at the eavesdropper is at least as large as that at the legitimate destination, does not seem to be immediately generalized to the scenario with fixed channel gains. 
\end{rem}
\begin{rem}
We note that our achievability results can be generalized for complex channel fading coefficients by applying Lemma 7 of \cite{MaddahAli:09} in our analysis of interference alignment. Also, our converse result for the case with full CSI can be generalized for complex channel fading coefficients in a straightforward manner. 
\end{rem}

%%%%%%%%%%%%%%%%%%%%%%%%%%%%%%%%%%%%%%%%%%%%%%%%%%%%
\section{Converse} \label{sec:converse}
For the Gaussian multiple access-wiretap channel, it is shown in Section 4.2.1 of \cite{MukherjeeXieUlukus:arxiv15} that there is no loss of secure d.o.f. if we consider the following deterministic model with integer-input and integer-output, instead of the model \eqref{eqn:model_y1}-\eqref{eqn:model_y2} in Section \ref{sec:model}: 
\begin{align}
Y_1(t)=\sum_{k=1}^2 \lfloor h_k(t)X_k(t)\rfloor,~ Y_2(t)=\sum_{k=1}^2 \lfloor g_k(t)X_k(t)\rfloor \label{eqn:det_output}
\end{align}
with the constraint 
\begin{align}
X_k\in \{0,1,\ldots,\lfloor \sqrt{P}\rfloor\}, k=1,2 \label{eqn:det_input} 
\end{align}
where $\lfloor \cdot \rfloor$ denotes the floor function.

Likewise, it can be shown that there is no loss of secure d.o.f. in considering the deterministic model \eqref{eqn:det_output} and \eqref{eqn:det_input} for the multiple access part in our Gaussian diamond-wiretap channel.\footnote{We omit a formal proof as it is identical to that in Section 4.2.1 of \cite{MukherjeeXieUlukus:arxiv15}.} Hence, in this section, let us assume that the multiple access part is given as \eqref{eqn:det_output} and \eqref{eqn:det_input}.  In this section, $c_i$'s for $i=1,2,3,\ldots$ are used to denote positive constants that do not depend on $n$ and $P$. We note that $\mathbf{h}^n, \mathbf{g}^n$ are known to the legitimate destination and the eavesdropper for the case with full CSI. For the case with no eavesdropper's CSI, we assume $\mathbf{g}^n$ in addition to $\mathbf{h}^n$ is available at the legitimate destination, which only possibly increases the secure d.o.f. Hence,  $\mathbf{h}^n, \mathbf{g}^n$  are conditioned in every entropy and mutual information terms in this section, but are omitted for notational convenience.

\subsection{Proof for the converse part of Theorem \ref{thm:CSI}} \label{subsec:conv}
From the cut-set bound, we can easily obtain
\begin{align}
d_s &\leq \min\{\alpha_1+\alpha_2, 1\}. \label{eqn:cut_dof}
\end{align}
Hence, it remains to show $d_s\leq \frac{\alpha_2+1}{2}.$
By applying the Fano's inequality, we have
\begin{align}
nR &\leq I(W;Y_1^n) + nc_2\cr
&\overset{(a)}{\leq} I(W;Y_1^n) - I(W;Y_2^n) + nc_3\cr
&\leq I(W;Y_1^n,Y_2^n)- I(W;Y_2^n) + nc_3\cr
& = I(W;Y_1^n|Y_2^n) + nc_3\cr
& \leq H(Y_1^n|Y_2^n)+ nc_3\cr
& = H(Y_1^n,Y_2^n) - H(Y_2^n) + nc_3 \cr
& \leq H(X_1^n,X_2^n,Y_1^n,Y_2^n)- H(Y_2^n) + nc_3\cr
&\leq H(X_1^n,X_2^n)+ H(Y_1^n,Y_2^n|X_1^n,X_2^n) - H(Y_2^n) + nc_3\cr
&\overset{(b)}{=} H(X_1^n,X_2^n) - H(Y_2^n) + nc_3,\label{eq3}
\end{align}
where $(a)$ is from the secrecy constraint and $(b)$ is because a deterministic model in \eqref{eqn:det_output} is assumed in this section. To bound $H(Y_2^n)$ in \eqref{eq3}, it follows that
\begin{align}
H(Y_2^n) &= H\big(\big\{\sum_{i = 1}^2 \lfloor g_i(t) X_i(t)\rfloor\big\}_{t=1}^n \big)\cr
&\geq H\big(\big\{\sum_{i = 1}^2 \lfloor g_i(t) X_i(t)\rfloor\big\}_{t=1}^n\big| X_2^n \big)\cr
& = H\big(\big\{ \lfloor g_1(t) X_1(t)\rfloor\big\}_{t=1}^n\big| X_2^n \big)\cr
& = H( X_1^n, \big\{ \lfloor g_1(t) X_1(t)\rfloor\big\}_{t=1}^n |X_2^n
) -H(X_1^n|\big\{ \lfloor g_1(t) X_1(t)\rfloor\big\}_{t=1}^n, X_2^n)\cr
& = H( X_1^n |X_2^n
) -H(X_1^n|\big\{ \lfloor g_1(t) X_1(t)\rfloor\big\}_{t=1}^n, X_2^n)\cr
&\geq  H(X_1^n |X_2^n) - \sum_{t=1}^n H(X_1(t)|\lfloor g_1(t) X_1(t)\rfloor) \cr
&\overset{(a)}{\geq} H(X_1^n |X_2^n) - nc_4, \label{eqn:y2}
\end{align}
where $(a)$ is from Lemma 2 of \cite{MukherjeeXieUlukus:arxiv15}.\footnote{We note that the constraint in Lemma 2 of \cite{MukherjeeXieUlukus:arxiv15} is satisfied under our channel model.} 

Now continuing \eqref{eq3} with \eqref{eqn:y2} substituted, we have
\begin{align}
nR &\leq H(X_2^n) + nc_5\cr
&\leq H(X_2^n, J_2)+nc_5\cr
&=H(J_2)+H(X_2^n|J_2)+nc_5. \label{eqBound1}
\end{align}

Note that the term $H(X_2^n|J_2) $ signifies the amount of individual randomness injected at relay $2$. Such individual randomness cannot be too large because of the reliability constraint at the receiver. To capture the trade-off between the rate $R$ and $H(X_2^n|J_2)$, we again start from the Fano's inequality to get
\begin{align}
nR &\leq I(W;Y_1^n) + nc_2\cr
&\leq I(J_1,J_2;Y_1^n) + nc_2\cr
& = H(Y_1^n) - H(Y_1^n|J_1,J_2) + nc_2.\label{eqM4}
\end{align}
For the term $H(Y_1^n|J_1,J_2)$, we have
\begin{align}
H(Y_1^n|J_1,J_2)&= H\big(\big\{\sum_{i = 1}^2 \lfloor h_i(t) X_i(t)\rfloor\big\}_{t=1}^n |J_1,J_2\big)\cr
&\geq H\big(\big\{\sum_{i = 1}^2 \lfloor h_i(t) X_i(t)\rfloor\big\}_{t=1}^n |J_1,J_2,X_1^n\big)\cr
&=  H\big(\big\{ \lfloor h_2(t) X_2(t)\rfloor\big\}_{t=1}^n |J_1,J_2,X_1^n\big)\cr
&=H( X_2^n, \big\{ \lfloor h_2(t) X_2(t)\rfloor\big\}_{t=1}^n  |J_1,J_2, X_1^n)-H(X_2^n|\big\{ \lfloor h_2(t) X_2(t)\rfloor\big\}_{t=1}^n ,J_1,J_2, X_1^n)\cr
&=H( X_2^n  |J_1,J_2, X_1^n)-H(X_2^n|\big\{ \lfloor h_2(t) X_2(t)\rfloor\big\}_{t=1}^n ,J_1,J_2, X_1^n)\cr
&\geq H( X_2^n  |J_1,J_2, X_1^n)-\sum_{t=1}^nH(X_2(t)|\lfloor h_2(t) X_2(t)\rfloor)\cr
&\overset{(a)}{\geq} H( X_2^n|J_1,J_2, X_1^n)-nc_6 \cr
&\overset{(b)}{=} H( X_2^n|J_2)-nc_6, \label{eqM2}
\end{align}
where $(a)$ is from Lemma 2 in \cite{MukherjeeXieUlukus:arxiv15} and $(b)$ is due to the Markov chain $X_2^n - J_2 - (X_1^n, J_1)$. Therefore, by substituting \eqref{eqM2} in \eqref{eqM4}, we  obtain
\begin{align}
nR \leq H(Y_1^n) - H(X_2^n|J_2)+nc_7.\label{eqM3} 
\end{align}
Combining \eqref{eqBound1} and \eqref{eqM3}, we have
\begin{align*}
2nR &\leq H(J_2) + H(Y_1^n)+nc_8. 
\end{align*}
Hence, we have
\begin{align*}
R& \leq \frac{1}{2}\left(\frac{1}{2}\log{P} + C_2\right) +  c_9,
\end{align*}
and, in turn, 
\begin{align*}
d_s \leq \frac{1}{2}(1 + \alpha_2).
\end{align*}
Combining with \eqref{eqn:cut_dof}, we finish the proof for the converse part of Theorem \ref{thm:CSI}.\endproof

\subsection{Proof for the converse part of Theorem \ref{thm:noCSI}} \label{subsec:conv_no}
Note that \eqref{eq3} continue to hold for the case with no eavesdropper's CSI. Continuing with \eqref{eq3}, it follows that
\begin{align}
nR &\leq H(X_1^n, X_2^n)-H(Y_2^n)+nc_{10}\cr
&\leq H(X_1^n, X_2^n, J_1, J_2) - H(Y_2^n)+ nc_{10}\cr
& = H(J_1, J_2) + H(X_1^n, X_2^n| J_1, J_2) - H(Y_2^n)+ nc_{10}\cr
&\leq H(J_1) + H(J_2) + H(X_1^n| J_1) + H(X_2^n|J_2) - H(Y_2^n)+nc_{10}.\label{eq:up1}
\end{align}
By applying similar steps as those to derive \eqref{eqM3}, we can obtain
\begin{align}
nR &\leq H(Y_1^n) - H(X_k^n|J_k)+nc_{11}\label{eq1},~ k=1,2.
\end{align}

Continuing with \eqref{eq:up1} substituted by \eqref{eq1} for $k=1,2$, we have
\begin{align*}
3nR \leq H(J_1)+H(J_2) + 2H(Y_1^n) - H(Y_2^n)+nc_{12}
\end{align*}
For the case with no eavesdropper's CSI, it is shown in Section 5 of \cite{DavoodiJafar:14} that the difference $H(Y_1^n) - H(Y_2^n)$ can not be larger than $n\cdot o(\log{P})$.\footnote{The channel assumption in \cite{DavoodiJafar:14}  is satisfied under our channel model.} Therefore, we have
\begin{align*}
3nR &\leq H(J_1)+H(J_2) + H(Y_1^n) + nc_{12}+n\cdot o(\log{P})
\end{align*}
which derives that
\begin{align*}
d_s \leq \frac{\alpha_1+\alpha_2 + 1}{3}.
\end{align*}
Since the bound on $d_s$ for the case with full CSI continues to hold for the case with no eavesdropper's CSI, we finish the proof for the converse part of Theorem \ref{thm:noCSI}. 
\endproof

%%%%%%%%%%%%%%%%%%%%%%%%%%%%%%%%%%%%%%%%%%%%%%%%%%%%%%%%%%%%%%%%%%%%%
\section{Achievability} \label{sec:achievability}
\begin{figure}[t]
 \centering\subfigure[]
  {\includegraphics[width=137mm]{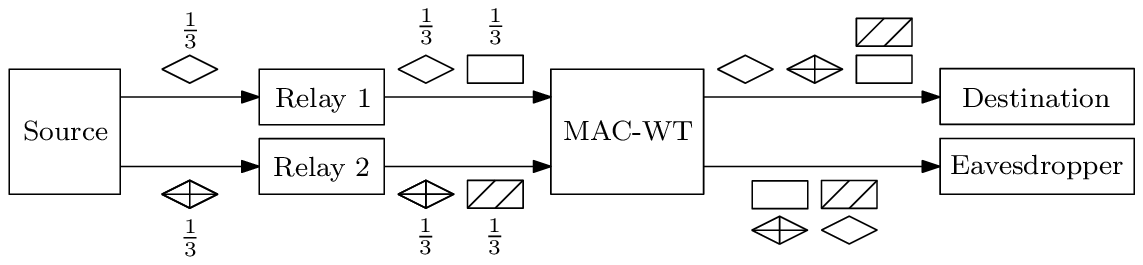}}
   \subfigure[]
  {\includegraphics[width=130mm]{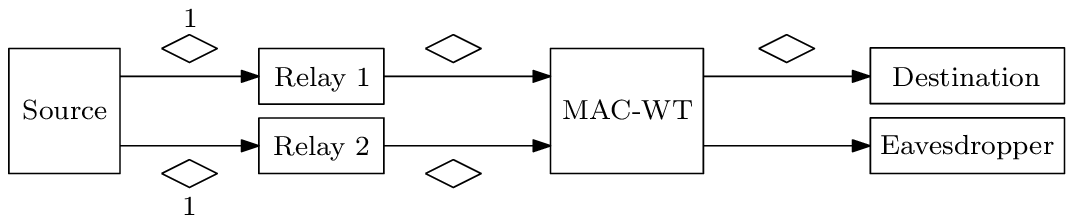}}
     \subfigure[]
  {\includegraphics[width=137mm]{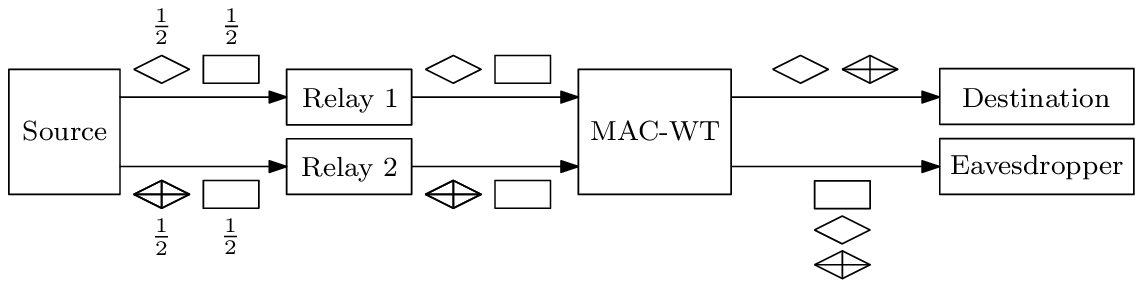}}
   \caption{Schemes for the case with full CSI:  (a) Incorporation of cooperative jamming, (b) Incorporation of message-beamforming, and (c) Simultaneous alignment and beamforming. Diamond shapes and rectangular shapes illustrate (partial) messages and noises, respectively, and the number above or below each shape represents its corresponding d.o.f. Same shapes with same patterns mean the same information. Otherwise, different shapes and/or different patterns represent independent informations. } \label{fig:with}
  \end{figure}

The direct parts of Theorems \ref{thm:CSI} and \ref{thm:noCSI} are proved by first identifying a few key constituent schemes and then time-sharing among them appropriately. Let us first provide a high-level description of those schemes and then give a detailed one. First, the following three schemes require the eavesdropper's CSI at the relays and the legitimate destination, whose operations are illustrated in Fig. \ref{fig:with}. 
\begin{itemize}
\item {[Scheme 1 achieving $(\alpha_1,\alpha_2,d_s)=(\frac{1}{3},\frac{1}{3},\frac{2}{3})$.  Incorporation of cooperative jamming \cite{XieUlukus:14}}]: \\ 
The message with d.o.f. $\frac{2}{3}$ is split into two independent partial messages each with d.o.f. $\frac{1}{3}$. The source sends a partial message to each relay in a way that each relay has a different partial message, which requires $\alpha_1=\alpha_2=\frac{1}{3}$. Then, the relays operate according to the cooperative jamming scheme \cite{XieUlukus:14}  for the Gaussian multiple access-wiretap channel, which is briefly explained in the following. Each relay sends independent partial message (d.o.f. $\frac{1}{3}$) together with its own noise (d.o.f. $\frac{1}{3}$) in a way that the noise signals are aligned at the legitimate destination and a  partial message signal sent from a relay is aligned with and is perfectly masked by the noise signal sent from the other relay at the eavesdropper.  

\item {[Scheme 2 achieving $(\alpha_1,\alpha_2,d_s)=(1,1,1)$. Incorporation of message-beamforming \cite{KhistiWornell:10,LiuLiuPoorShamai:10}}]: \\ The source sends the message with d.o.f. 1 to both the relays, which requires $\alpha_1=\alpha_2=1$. Both the relays send the message cooperatively in a way that the message signals are beam-formed in the null space of the eavesdropper's channel. 

\item {[Scheme 3 achieving $(\alpha_1,\alpha_2,d_s)=(1,1,1)$. Simultaneous alignment and beamforming (S-AB)]:} \\
The message with d.o.f. $1$ is split into two independent partial messages each with d.o.f. $\frac{1}{2}$. The source sends a partial message together with a common noise with d.o.f. $\frac{1}{2}$ to each relay, which requires $\alpha_1=\alpha_2=1$. Then, each relay sends independent partial message (d.o.f. $\frac{1}{2}$) and common noise (d.o.f. $\frac{1}{2}$) in a way that the common noise signals are beam-formed in the null space of the legitimate destination's channel and the partial message signals are aligned with and are perfectly masked by the common noise signal at the eavesdropper. Although this scheme achieves the same d.o.f. tuple as for Scheme 2, it outperforms Scheme 2 for more than two relays as remarked in Section~\ref{sec:multi}.

\end{itemize}

\begin{figure}
  \centering 
 \subfigure[]
  {\includegraphics[width=130mm]{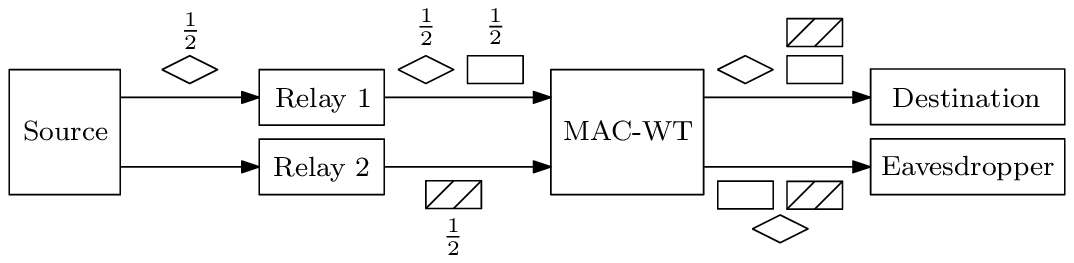}}
   \subfigure[]
  {\includegraphics[width=130mm]{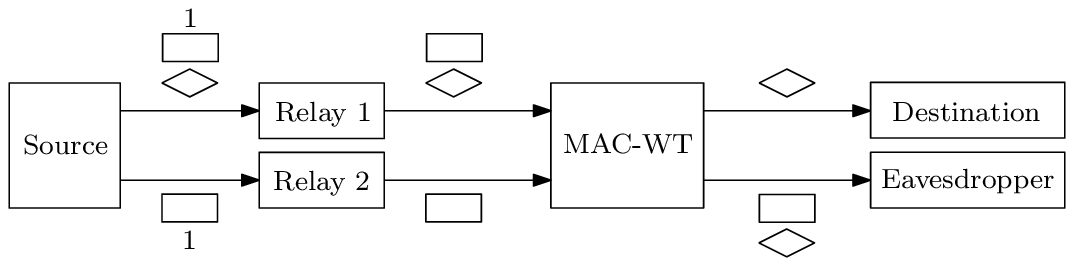}}
   \caption{Schemes for the case with no eavesdropper's CSI:  (a)  Incorporation of blind cooperative jamming and (b)  Computation for jamming. Similarly as in Fig. \ref{fig:with}, diamond shapes and rectangular shapes represent (partial) messages and noises, respectively, with the number above or below each shape corresponding to its d.o.f. Same shapes with same patterns represent the same information, and otherwise independent informations.   
     } \label{fig:no_csi}

\end{figure}

Next, the following two schemes operate with no eavesdropper's CSI at the relays and the legitimate destination, which are illustrated in Fig. \ref{fig:no_csi}.
\begin{itemize}
\item  {[Scheme 4 achieving $(\alpha_1,\alpha_2,d_s)=(\frac{1}{2},0,\frac{1}{2})$. Incorporation of blind cooperative jamming \cite{MukherjeeXieUlukus:arxiv15}]: } \\ The source sends the message with d.o.f. $\frac{1}{2}$ only to relay 1, which requires $\alpha_1=\frac{1}{2}$. Then, the relays operate according to the blind cooperative jamming scheme \cite{MukherjeeXieUlukus:arxiv15} for the wiretap channel with helpers. Relay 1 sends the  message (d.o.f. $\frac{1}{2}$) together with its own noise (d.o.f. $\frac{1}{2}$) and relay 2 sends its own noise (d.o.f. $\frac{1}{2}$) in a way that the noise signals are aligned at the legitimate destination. Since the noise signals occupy the entire space at the eavesdropper, the messages can be shown to be secure.
  
\item {[Scheme 5 achieving $(\alpha_1,\alpha_2,d_s)=(1,1,1)$. Computation for jamming (CoJ)]:}\\
The source adds a noise sequence with d.o.f. 1 to the message codeword with d.o.f. 1  and sends the resultant sequence, which also has d.o.f. 1, to relay 1. To relay 2, the source sends the noise sequence used for the addition. This requires $\alpha_1=\alpha_2=1$. Then, relays 1 and 2 send what they have received in a way that the common noise signals are canceled out at the legitimate destination. Because the common noise signals occupy the entire space at the eavesdropper, the message can be shown to be secure. 
\end{itemize}

To show the achievability part of Theorem \ref{thm:CSI}, we perform time-sharing among Scheme 1, Scheme 4, and any of Schemes 2, 3, and 5. For the achievability part of Theorem \ref{thm:noCSI}, we time-share between Schemes 4 and 5. Because Schemes 1, 2, and 4 are straightforward extensions of the previously proposed schemes  in \cite{XieUlukus:14,KhistiWornell:10,LiuLiuPoorShamai:10,MukherjeeXieUlukus:arxiv15}, we give a detailed description only for Schemes 3 and 5. To that end, we first present some achievability results  for the Gaussian multiple access-wiretap channel, which corresponds to the multiple access part of our model where each relay acts as a source having its own message. In the Gaussian multiple access-wiretap channel, source $k=1,2$ wishes to send message $W_k$ of rate $R_k$ to the legitimate destination  while keeping it secret from the eavesdropper. A secrecy rate tuple $(R_1, R_2)$ is said to be achievable if there exists a sequence of codes with block length $n$ such that $\lim_{n\rightarrow \infty }P(\hat{W}_1\neq W_1 \mbox{ or } \hat{W}_2\neq W_2)=0$ and $\lim_{n\rightarrow \infty}\frac{1}{n}I(W_1, W_2;Y_2^n|\mathbf{h}^n, \mathbf{g}^n)=0$. The following two theorems give achievable secrecy rate regions for the Gaussian multiple access-wiretap channel for the case with full CSI at the sources and the legitimate destination and for the case with no eavesdropper's CSI at the sources  and the legitimate destination, respectively. Since these theorems are direct consequences of the achievablility result in \cite{TekinYener:08general}, their proofs are omitted in this paper. 
\begin{thm}\label{thm:mac_1}
For the Gaussian multiple access-wiretap channel with full CSI at the sources and the legitimate destination, a secrecy rate tuple $(R_1, R_2)$ is achievable if 
\begin{align*} 
\sum_{k\in S} R_k \leq I(V_S;Y_1|V_{S^c},\mathbf{h},\mathbf{g})-I(V_S;Y_2|\mathbf{h},\mathbf{g})
\end{align*}
for all $S\subseteq [1:2]$ for some $\prod_{k\in[1:2]} p(v_k)p(x_k|v_k,\mathbf{h},\mathbf{g})$ such that $\E[X_k^2]\leq P$ for $k=1,2$.\footnote{In Theorems \ref{thm:mac_1} and \ref{thm:mac_2}, $\mathbf{h}=(h_1,h_2)$ and $\mathbf{g}=(g_1,g_2)$ denote the random channel fading coefficients generated by $f(h_1,h_2,g_1,g_2)$.  }
\end{thm}

\begin{thm}\label{thm:mac_2}
For the Gaussian multiple access-wiretap channel with no eavesdropper's CSI at the sources and the legitimate destination, a secrecy rate tuple $(R_1,R_2)$ is achievable if 
\begin{align*}
\sum_{k\in S} R_k \leq I(V_S;Y_1|V_{S^c},\mathbf{h})-I(V_S;Y_2|\mathbf{h},\mathbf{g})
\end{align*}
for all $S\subseteq [1:2]$ for some $\prod_{k\in[1:2]} p(v_k)p(x_k|v_k,\mathbf{h})$ such that $\E[X_k^2]\leq P$ for $k=1,2$.  
\end{thm}
\begin{rem}
Theorems \ref{thm:mac_1} and \ref{thm:mac_2} can be obtained from \cite{TekinYener:08general} by applying the technique of adding prefix channels introduced in \cite{CsiszarKorner:78}. We add
 prefix channel $p(x_k|v_k,\mathbf{h},\mathbf{g})$ for the case with full CSI and add prefix channel $p(x_k|v_k,\mathbf{h})$ for the case with no eavesdropper's CSI. 
\end{rem}

Now, let us describe Schemes 3 and 5.

\subsubsection*{Scheme 3 achieving $(\alpha_1,\alpha_2,d_s)=(1,1,1)$. Simultaneous alignment and beamforming scheme}
The message $W$ of rate $R$ is split into $W_1$ and $W_2$ each of which having rate $R/2$. Then, the source sends $W_k$ to relay $k$ together with a common noise sequence $U^n$ generated in an i.i.d. manner according to $\mbox{Unif}[C(a,Q)]$, where 
\begin{align*}
C(a,Q)=a\{-Q,-Q+1,\cdots,0,\cdots,Q-1,Q\}
\end{align*}
for some positive real number $a$ and positive integer $Q$ which will be specified later. This transmission from the source to the relays imposes the following constraints:
\begin{align}
\frac{R}{2}+\log (2Q+1)&\leq C_1 \label{eqn:cn_1}\\
\frac{R}{2}+\log (2Q+1)&\leq C_2 \label{eqn:cn_2}.
\end{align}

Now, we apply Theorem \ref{thm:mac_1} for the multiple access part with the following choices of $R_1$, $R_2$, and $p(v_1)p(v_2)p(x_1|v_1, \mathbf{h}, \mathbf{g})p(x_2|v_2,\mathbf{h}, \mathbf{g})$: 
\begin{align*}
R_1=R/2,&~R_2=R/2\\
V_1\sim\mbox{Unif}[C(a,Q)],&~ V_2\sim\mbox{Unif}[C(a,Q)] \\
X_1= \left(h_2 - \frac{g_2}{g_1}h_1\right){V}_1 + h_2{U},&~X_2= \left(\frac{g_1}{g_2}h_2 - h_1\right){V}_2 -h_1{U}.
 \end{align*}
Then, the channel outputs at the legitimate destination and the eavesdropper are given as 
 \begin{align*}
Y_1 &= \left(h_1h_2 - \frac{g_2}{g_1}h_1^2\right){V}_1 + \left(\frac{g_1}{g_2}h_2^2 - h_1h_2\right){V}_2 + N_1\\
Y_2 &= (g_1h_2 - g_2h_1)({V}_1 + {V}_2 + {U}) + N_2,
\end{align*}
respectively. According to Theorem \ref{thm:mac_1}, the secrecy rate of $R$ is achievable if 
\begin{align}
R &\leq I(V_1,V_2;Y_1|\mathbf{h}, \mathbf{g}) - I(V_1,V_2;Y_2|\mathbf{h}, \mathbf{g}) \label{eqrate} \\
\frac{R}{2} &\leq I(V_1;Y_1|V_2, \mathbf{h}, \mathbf{g}) - I(V_1;Y_2|\mathbf{h}, \mathbf{g}) \label{eqrate1} \\
\frac{R}{2} &\leq I(V_2;Y_1|V_1,\mathbf{h}, \mathbf{g}) - I(V_2;Y_2|\mathbf{h}, \mathbf{g}) \label{eqrate2}
\end{align} 
are satisfied. 
  
Let us bound the first term in the RHS of \eqref{eqrate}. The constellation at the legitimate destination consists of $(2Q + 1)^2$ points and the minimum distance $d_{\min}$ of which can be bounded using the Khintchine-Groshev theorem of Diophantine approximation \cite{MotahariOveisGharanMaddahAliKhandani:14} as follows: for any $\delta > 0$, there exists a constant $k_{\delta}$ such that
\begin{align}
d_{\min} \geq \frac{ak_{\delta}}{Q^{1 + \delta}} \label{eqn:d_min}
\end{align}
for almost all channel fading coefficients except a set of Lebesque measure zero. Since the probability that a realization of channel fading coefficients does not satisfy \eqref{eqn:d_min} is negligible, for the sake of brevity, let us assume that channel fading coefficients satisfy \eqref{eqn:d_min} in the subsequent analysis.
 
Let $(\hat{V}_1, \hat{V}_2)$ denote the estimate of $(V_1,V_2)$ which is chosen as the closest point to $Y_1$ in the constellation.  Then, we have 
 \begin{align*}
P((\hat{V}_1, \hat{V}_2)\neq (V_1,V_2)) &\leq \exp\left(-\frac{d_{\min}^2}{8}\right)\\
    &\leq \exp\left(-\frac{a^2k_{\delta}^2}{8Q^{2(1 + \delta) }}\right).
 \end{align*}
 By choosing $Q = P^{\frac{1 - \delta}{2(2 + \delta)}}$ and $a = \frac{\gamma P^{1/2}}{Q}$ for some $\gamma>0$, we  have
 \begin{align*}
P((\hat{V}_1, \hat{V}_2)\neq (V_1,V_2)) &\leq \exp\left(-\frac{\gamma^2k_{\delta}^2P^{\delta}}{8}\right).
 \end{align*}
To meet the power constraints, we choose $\gamma$ as follows:
\begin{align*}
\gamma = \frac{1}{\sqrt{5}L^3}\leq \min\left\{\frac{1}{\sqrt{(h_2 - \frac{g_2}{g_1}h_1)^2 + h_2^2}},\frac{1}{\sqrt{(\frac{g_1}{g_2}h_2 - h_1)^2 + h_1^2}}\right\}.
\end{align*}
 According to the Fano's inequality, it follows that
\begin{align*}
H(V_1,V_2|Y_1,\mathbf{h}, \mathbf{g}) &\leq  H(V_1,V_2|\hat{V}_1,\hat{V}_2)\\
&\leq 1 + P((\hat{V}_1, \hat{V}_2)\neq (V_1,V_2)) \log(|(V_1,V_2)| - 1)\\
& \leq 1 + \exp\left(-\frac{k_{\delta}^2P^{\delta}}{40L^6}\right)\log{(2Q + 1)^2}\notag\\
& = o(\log{P}).
\end{align*}
Therefore, the first term in the RHS of \eqref{eqrate} can be bounded as
 \begin{align}
 I(V_1,V_2;Y_1|\mathbf{h}, \mathbf{g})& = H(V_1,V_2|\mathbf{h}, \mathbf{g}) - H(V_1,V_2|Y_1,\mathbf{h}, \mathbf{g})\cr
  &\geq \log{(2Q + 1)^2} - o(\log P)\cr
  &= \frac{1 - \delta}{2 + \delta}\log P - o(\log{P})\label{eqterm1}.
 \end{align}

For the second term in the RHS of \eqref{eqrate}, it follows that
 \begin{align}
I(V_1,V_2;Y_2|\mathbf{h}, \mathbf{g})&\overset{(a)}{\leq} I(V_1,V_2;(g_1h_2 - g_2h_1)(V_1 + V_2 + U)|\mathbf{h}, \mathbf{g})\cr
&\overset{(b)}{=} I(V_1,V_2; V_1 + V_2 + U) \cr
 &= H(V_1 + V_2 + U) - H(U) \cr
 &\leq \log(6Q + 1) - \log(2Q+1)\cr
 & = \log\left(\frac{6Q + 1}{2Q + 1}\right)\cr
 & = o(\log{P})\label{eqterm2},
 \end{align}
where $(a)$  is due to the Markov chain $(V_1,V_2) - ((g_1h_2 - g_2h_1)(V_1 + V_2 + U),\mathbf{h}, \mathbf{g}) - Y_2$ and $(b)$ is because $P(g_1h_2 - g_2h_1=0)=0$ for our channel model.

Next, for the first term in the RHS of \eqref{eqrate1}, we have 
\begin{align}
I(V_1;Y_1|V_2, \mathbf{h}, \mathbf{g})&= I(V_1;Y_1'| \mathbf{h}, \mathbf{g})\cr
&=H(V_1)-H(V_1|Y_1',\mathbf{h}, \mathbf{g})\cr
&\overset{(a)}{=}H(V_1)-h(N_1|Y_1',\mathbf{h}, \mathbf{g})\cr
&\geq H(V_1)-h(N_1) \cr
&=\log (2Q+1)-\frac{1}{2}\log 2\pi e \cr
&\geq\frac{1-\delta}{2(2+\delta)}\log P-o(\log P)  \label{eqn:no_inter}
\end{align}
for $Y_1'\triangleq h_{\mathrm{eff}}V_1+N_1$ and $h_{\mathrm{eff}}\triangleq h_1h_2 - \frac{g_2}{g_1}h_1^2$, where $(a)$ is because  $P(h_{\mathrm{eff}}= 0)=0$ for our channel model  and for given $Y_1', \mathbf{h}, \mathbf{g}$ with $h_{\mathrm{eff}}\neq 0$, $V$ and $N_1$ have a one-to-one relationship. For the second term in the RHS of \eqref{eqrate1}, we have 
\begin{align}
I(V_1;Y_2|\mathbf{h}, \mathbf{g})&\leq I(V_1,V_2;Y_2|\mathbf{h}, \mathbf{g}) \cr
 &\overset{(a)}{\leq} o(\log{P}), 
\end{align}
where $(a)$ is from \eqref{eqterm2}.

Similarly, for the terms in the RHS of  \eqref{eqrate2}, we can show 
\begin{align}
I(V_2;Y_1|V_1, \mathbf{h}, \mathbf{g})&\geq\frac{1-\delta}{2(2+\delta)}\log P-o(\log P)  \\
I(V_2;Y_2|\mathbf{h}, \mathbf{g})&\leq  o(\log{P}). \label{eqn:term_last}
\end{align}

By substituting \eqref{eqrate}-\eqref{eqrate2} with \eqref{eqterm1}-\eqref{eqn:term_last} and then choosing $\delta$ sufficiently small, we have 
\begin{align}
R\leq \frac{1}{2}\log{P} - o(\log{P})\label{eqn:cn}
\end{align}
for the multiple access part. From  \eqref{eqn:cn_1}, \eqref{eqn:cn_2}, and \eqref{eqn:cn}, we conclude that $(\alpha_1,\alpha_2,d_s)=(1,1,1)$ is achievable. 

\subsubsection*{Scheme 5 achieving $(\alpha_1, \alpha_2, d_s)=(1,1,1)$. Computation for jamming scheme}
In this scheme, we wish to apply Theorem \ref{thm:mac_2} for the multiple access part with the following choices of $R_1$, $R_2$, and $p(v_1)p(v_2)p(x_1|v_1,\mathbf{h})p(x_2|v_2,\mathbf{h})$:
\begin{align*}
R_1=R,&~ R_2=0\\
V_1\sim\mbox{Unif}[C(a,Q)],&~ V_2=\emptyset \\
X_1= \frac{1}{h_1} ({V}_1+U),&~X_2= -\frac{1}{h_2}{U},
 \end{align*}
where $U\sim\mbox{Unif}[C(a,Q)]$, $C(a,Q)=a\{-Q,-Q+1,\cdots,0,\cdots,Q-1,Q\}$, $Q = P^{\frac{1 - \delta}{2}}$, and $a = \frac{1}{\sqrt{2}L} P^{\frac{\delta}{2}}$ for $\delta>0$. Note that the power constraints at the relays are satisfied since $\frac{1}{\sqrt{2}L}\leq \min\left\{\frac{|h_1|}{\sqrt{2}},|h_2|\right\}$.

  To that end, one naive approach is to let the source send the message $W$ to relay 1 and send a common noise sequence $U^n$ to both relays 1 and 2, which requires
\begin{align*}
R+\log (2Q+1) &\leq C_1\\
\log(2Q+1) &\leq C_2.
\end{align*}
However, there is a cleverer way to enable the aforementioned relay operations, in which the source \emph{computes} $V_1^n(W)+U^n$ and sends the sum to relay 1. To relay 2, the source sends $U^n$. This transmission from the source to the relays is possible if the following constraints are satisfied:
\begin{align}
\log (4Q+1) &\leq C_1 \label{eqn:cn_1_no}\\
\log(2Q+1) &\leq C_2. \label{eqn:cn_2_no}
\end{align}

Now, the channel outputs at the legitimate destination and the eavesdropper are given as 
 \begin{align*}
Y_1 &=  V_1+N_1\cr
Y_2 &= \frac{g_1}{h_1} V_1+\left(\frac{g_1}{h_1}-\frac{g_2}{h_2}\right)U+ N_2,
\end{align*}
respectively. According to Theorem \ref{thm:mac_2}, the following secrecy rate can be achieved.
\begin{align}\label{eqrate_no}
R \leq I(V_1;Y_1|\mathbf{h}) - I(V_1;Y_2|\mathbf{h}, \mathbf{g}).
\end{align} 
  
Let us bound the first term in the RHS of \eqref{eqrate_no}. We have 
\begin{align}
I(V_1;Y_1|\mathbf{h})&=I(V_1;V_1+N_1)\cr
&=H(V_1)-H(V_1|V_1+N_1)\cr
&=H(V_1)-h(N_1|V_1+N_1)\cr
&\geq H(V_1)-h(N_1)\cr
&=\log (2Q+1)-\frac{1}{2}\log (2\pi e)\cr
&\geq \frac{1-\delta}{2} \log P -o(\log P). \label{eqterm1_no}
\end{align}

For the second term in the RHS of \eqref{eqrate_no}, it follows that
 \begin{align}
I(V_1;Y_2|\mathbf{h}, \mathbf{g})&= I(V_1,U;Y_2|\mathbf{h}, \mathbf{g})-I(U;Y_2|V_1,\mathbf{h}, \mathbf{g})\cr
&\overset{(a)}{\leq} I(V_1,U;Y_2|\mathbf{h}, \mathbf{g})- \frac{1-\delta}{2}\log P+o(\log P) \cr 
&=h(Y_2|\mathbf{h}, \mathbf{g})-h(Y_2|V_1,U,\mathbf{h}, \mathbf{g})- \frac{1-\delta}{2}\log P+o(\log P) \cr
&=h(Y_2|\mathbf{h}, \mathbf{g})-h(N_2)-\frac{1-\delta}{2}\log P+o(\log P) \cr
&\overset{(b)}{\leq} \frac{1}{2}\log P-\frac{1}{2}\log 2\pi e -\frac{1-\delta}{2}\log P+o(\log P) \cr
&=\frac{\delta}{2}\log P+o(\log P), \label{eqnterm2_no}
\end{align}
where $(a)$ is by applying similar steps as those used for obtaining \eqref{eqn:no_inter} and $(b)$ is because all channel fading coefficients are assumed to be bounded away from zero and infinity. 
 
By substituting \eqref{eqrate_no} with \eqref{eqterm1_no} and \eqref{eqnterm2_no} and by choosing $\delta$ sufficiently small, we have
 \begin{align}
 R \leq \frac{1}{2}\log{P} + o(\log{P}) \label{eqn:cn_no}
 \end{align}
for the multiple access part. From \eqref{eqn:cn_1_no}, \eqref{eqn:cn_2_no}, and \eqref{eqn:cn_no}, we conclude that $(\alpha_1,\alpha_2,d_s)=(1,1,1)$ is achievable. 

Now, we are ready to prove the achievability parts of Theorems \ref{thm:CSI} and \ref{thm:noCSI}. 
\subsection{Proof for the achievability part of Theorem \ref{thm:CSI}}
Note that $\alpha_1\geq \alpha_2$ without loss of generality in our model. First, consider the case where the minimum of \eqref{eqn:CSI} is equal to $\alpha_1+\alpha_2$, which implies $2\alpha_1+\alpha_2\leq 1$.
We use time-sharing technique as follows: use Scheme 1 for $3\alpha_2$ fraction of time, use Scheme 4 for $2(\alpha_1-\alpha_2)$ fraction of time, and keep silent for the remaining fraction.\footnote{Note that $3\alpha_2+2(\alpha_1-\alpha_2)=2\alpha_1+\alpha_2\leq 1$.} Then, it can be easily shown  $d_s=\alpha_1+\alpha_2$ is achievable. Next, consider the case where the minimum of \eqref{eqn:CSI} is given as  $\frac{1}{2}(1+\alpha_2)$. If $\alpha_2\leq \frac{1}{3}$, by using Scheme 1 for $3\alpha_2$ fraction of time and using Scheme 4 for $1-3\alpha_2$ fraction of time, $\frac{1}{2}(1+\alpha_2)$ is achievable. If $\frac{1}{3}<\alpha_2\leq 1 $, by using Scheme 1 for  $\frac{3}{2}(1-\alpha_2)$ fraction of time and using any of Schemes 2, 3, and 5 for the remaining fraction of time, $\frac{1}{2}(1+\alpha_2)$ is achievable. Finally, consider the case where the minimum of \eqref{eqn:CSI} is 1, which implies $\alpha_1\geq 1$ and $\alpha_2\geq 1$. By using any of Schemes 2, 3, and 5, $d_s=1$ is trivially achievable. \endproof

\subsection{Proof for the achievability part of Theorem \ref{thm:noCSI}}
 We note that a variant of Scheme 4 where the roles of relays 1 and 2 are swapped  can achieve $(\alpha_1,\alpha_2,d_s)=(0,\frac{1}{2},\frac{1}{2})$, and let us call this scheme as Scheme 4$^*$. 
First, consider the case where the minimum of \eqref{eqn:CSI_no} is equal to $\alpha_1+\alpha_2$, which implies $2\alpha_1+2\alpha_2\leq 1$. By using Scheme 4 for $2\alpha_1$ fraction of time and Scheme 4$^*$ for $2\alpha_2$ fraction of time and keeping silent for the remaining fraction, $d_s=\alpha_1+\alpha_2$ can be shown to be achievable. Next, consider the case where the minimum of \eqref{eqn:CSI_no} is given as  $\frac{1}{3}(1+\alpha_1+\alpha_2)$. By using Scheme 4 for $\frac{2(\alpha_1-2\alpha_2+1)}{3}$ fraction, Scheme 4$^*$ for $\frac{2(\alpha_2-2\alpha_1+1)}{3}$ fraction, and Scheme 5 for the remaining fraction of time, it can be shown that $d_s=\frac{1}{3}(1+\alpha_1+\alpha_2)$ is achievable. Now, consider  the case where the minimum of \eqref{eqn:CSI_no} is given as  $\frac{1}{2}(1+\alpha_2)$. We use Scheme 4 for $(1-\alpha_2)$ fraction of time and use Scheme 5 for $\alpha_2$ fraction of time, which achieves $d_s=\frac{1}{2}(1+\alpha_2)$. 
Finally, consider the case where the minimum of \eqref{eqn:CSI_no} is 1, which implies $\alpha_1\geq 1$ and $\alpha_2\geq 1$. By using Scheme 5, $d_s=1$ is trivially achievable. \endproof

\section{Generalization to $M$ relays}\label{sec:multi}
In this section, we consider a generalized Gaussian diamond-wiretap channel where there are arbitrary number of relays. Assume that there are $M\geq 2$ relays with transmit power constraint of $P$. For the broadcast part,  the source is connected to  $M$ relays through orthogonal links, where the link capacity to relay $k=1,\ldots, M$ is $C_k$ such  that $\lim_{P\rightarrow \infty} \frac{C_k}{\frac{1}{2}\log P}=\alpha_k$. For the multiple access part, the channel outputs $Y_1(t)$ and $Y_2(t)$ at time $t$ at the legitimate destination and the eavesdropper, respectively, are given as 
\begin{align}
Y_1(t) = \sum_{k = 1}^M h_k(t) X_k(t) + N_1(t) \label{eqn:y1_M}\\
Y_2(t) = \sum_{k = 1}^M g_k(t) X_k(t) + N_2(t),\label{eqn:y2_M}
\end{align}
in which $X_k(t)$ is the channel input at relay $k$, $h_k(t)$'s and $g_k(t)$'s are channel fading coefficients to the legitimate destination and the eavesdropper, respectively, and $N_1(t)$ and $N_2(t)$ are independent Gaussian noise with zero mean and unit variance at the legitimate destination and the eavesdropper, respectively,  at time $t$. Similarly as in Section \ref{sec:model}, we assume fast fading\footnote{Similarly as for the two-relay case in Section \ref{sec:model}, we assume that the channel fading coefficients are generated from a real-valued joint density function whose all joint and conditional density functions are bounded and whose support set does not contain zero and infinity.}, no CSI at the source, and full CSI at the eavesdropper, and consider the two cases regarding the availability of CSI at the relays and the legitimate destination, i.e., the case with full CSI and the case with no eavesdropper's CSI. 
A secrecy code, secrecy capacity, and secure d.o.f. are defined by a straightforward generalization from Section \ref{sec:model}. For the brevity of the results, we focus on the symmetric case with $C_1=\cdots=C_M=C$, which implies  $\alpha_1=\cdots=\alpha_M=\alpha$. 

The following two theorems present our results on the secure d.o.f. for the case with full CSI and for the case with no eavesdropper's CSI, respectively. 

\begin{thm}\label{thm2}
For the generalized Gaussian diamond-wiretap channel with $M\geq 2$ relays with full CSI at the relays and the legitimate destination, the secure d.o.f.  satisfies
\begin{align*}
d_{s,-} \leq d_s \leq d_{s,+},
\end{align*}
where
\begin{align*}
d_{s,+} &= \min\left\{M\alpha,\frac{M - 1}{M}(1 + \alpha),1\right\},\\
d_{s,-} &= \min\left\{M\alpha, \frac{2M(M - 1) + M^2\alpha}{2M^2 - M + 2}, 1\right\}.
\end{align*}
\end{thm}

\begin{thm} \label{thm2_no}
For the generalized Gaussian diamond-wiretap channel with $M\geq 2$ relays with no eavesdropper's CSI at the relays and the legitimate destination, the secure d.o.f.  is equal to 
\begin{align*}
d_{s} &= \min\left\{M\alpha,\frac{M\alpha + M - 1}{M + 1},1\right\}.
\end{align*}
\end{thm}

\begin{figure}
  \centering
  \includegraphics[width=140mm]{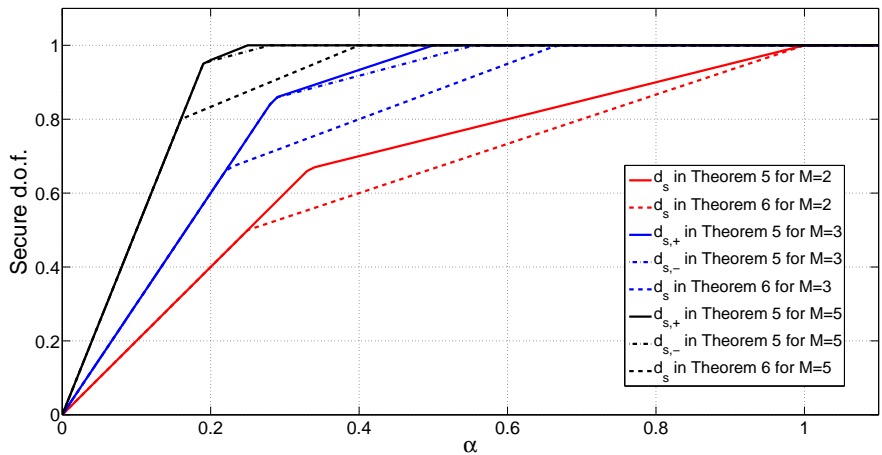}\\
  \caption{Secure d.o.f. of the generalized Gaussian diamond-wiretap channel with $M$ relays}\label{fig:multiple}
\end{figure}

In Fig. \ref{fig:multiple}, the results in Theorems \ref{thm2} and \ref{thm2_no} are illustrated for $M=2,3,5$. For the case with full CSI with $M>2$, there exists a gap between the lower and upper bounds on the secure d.o.f., which decreases as $M$ increases. 
We note that up to the threshold value $\frac{M-1}{M(M-1)+1)}$ (resp.,  $\frac{M-1}{M^2}$)  of $\alpha$ for the case with full CSI (resp., for the case with no eavesdropper's CSI), the secure d.o.f is linear in $M$ and $\alpha$. In this regime, the broadcast part becomes the bottleneck and hence it is optimal to send independent partial messages to the relays and  to incorporate the cooperative jamming scheme \cite{XieUlukus:14} (resp., the blind cooperative jamming scheme \cite{MukherjeeXieUlukus:arxiv15}) for the multiple access part.  After this threshold value of $\alpha$, the source needs to send some common information to the relays to achieve a higher secure d.o.f. and hence the slope of secure d.o.f. in $\alpha$ becomes lower. 
If $\alpha$ is  $\frac{1}{M}+\frac{2}{M^2}$ for the case with full CSI  (resp., $\frac{2}{M}$ for the case with no eavesdropper's CSI), one secure d.o.f. can be achieved by generalizing the S-AB scheme (resp., the CoJ scheme) in Section \ref{sec:achievability}. We note that there is  a gap between the secure d.o.f.'s with and without eavesdropper's CSI for $\alpha\in \left(\frac{M-1}{M^2}, \frac{2}{M}\right)$, and there is no loss in secure d.o.f. for other ranges of $\alpha$. 

Theorems \ref{thm2} and \ref{thm2_no}  can be proved by generalizing the proof techniques in Sections 
\ref{sec:converse} and \ref{sec:achievability}. In the following, we provide brief proofs for these theorems. 

\subsection{Converse}
Similarly as in Section \ref{sec:converse}, there is no loss of secure d.o.f. in considering  the following deterministic model with integer-input and integer-output for the multiple access part, instead of the original channel \eqref{eqn:y1_M} and \eqref{eqn:y2_M}: 
\begin{align}
Y_1(t)=\sum_{k=1}^M \lfloor h_k(t)X_k(t)\rfloor,~ Y_2(t)=\sum_{k=1}^M \lfloor g_k(t)X_k(t)\rfloor \label{eqn:ch_M}
\end{align}
with the constraint 
\begin{align}
X_k\in \{0,1,\ldots,\lfloor \sqrt{P}\rfloor\}, k=1,\ldots, M. \label{eqn:pw_M}
\end{align}
Hence, in this subsection, let us assume that the multiple access part is given as \eqref{eqn:ch_M} and \eqref{eqn:pw_M}. 
In addition, the channel fading coefficients are conditioned in every entropy and mutual information terms in this subsection due to the same reason as in Section \ref{sec:converse}, but are omitted for notational convenience. $c_i'$'s for $i=1,2,3,\ldots$ are used to denote positive constants that do not depend on $n$ and $P$.

\subsubsection{Proof for the converse part of Theorem \ref{thm2}}
We generalize the converse proof technique in Section \ref{subsec:conv} for multiple relays. We can obtain the following inequality by applying similar techniques used to obtain \eqref{eqBound1}:  
\begin{align}
nR &\leq  \sum_{k = 2}^M H(J_k) + \sum_{k = 2}^M H(X_k^n|J_k)+ nc_1'. \label{eq:conv_f_multi1}
\end{align}
On the other hand, we can generalize \eqref{eqM3} for multiple relays as follows: 
\begin{align}
nR \leq H(Y_1^n) - H(X_k^n|J_k)+nc_2', ~k = 2,\cdots,M.\label{eq:conv_f_multi2}
\end{align}
By combining \eqref{eq:conv_f_multi1} and \eqref{eq:conv_f_multi2}, we have
\begin{align*}
MnR &\leq \sum_{k = 2}^M H(J_k) + (M - 1)H(Y_1^n)+nc_3' \\
&\leq (M - 1)nC + (M - 1)H(Y_1^n)+nc_3'.
\end{align*}
It follows that
\begin{align*}
R& \leq \frac{M - 1}{M}\left(\frac{1}{2}\log{P} + C\right) + c_4'
\end{align*}
or
\begin{align*}
d_s \leq \frac{M - 1}{M}(1 + \alpha).
\end{align*}
Together with the following bound from the cutset bound, this completes the proof,
\begin{align}
d_s\leq \min \{M\alpha, 1\}. \label{eqn:cutset_multi}
\end{align}
 \endproof

\subsubsection{Proof for the converse part of Theorem \ref{thm2_no}}
We extend the converse proof technique in Section \ref{subsec:conv_no} for multiple relays. First, we can generalize \eqref{eq:up1} for multiple relays as follows:  
\begin{align}
nR \leq \sum_{k = 1}^M H(J_k) + \sum_{k = 1}^M H(X_k^n|J_k) - H(Y_2^n)+nc_5'.\label{eq:step1Mrelay}
\end{align}
Next, the following inequality can be obtained by applying similar techniques used in deriving \eqref{eq1}: 
\begin{align}
nR \leq H(Y_1^n) - H(X_k^n|J_k)+nc_6'\label{eq:step2Mrelay}, ~k=1,\ldots, M
\end{align}
Combining \eqref{eq:step1Mrelay} and \eqref{eq:step2Mrelay}, we have
\begin{align*}
(M + 1)nR &\leq \sum_{k = 1}^M H(J_k) + MH(Y_1^n) - H(Y_2^n)+nc_7'\\
&\leq MnC + (M-1)H(Y_1^n) +H(Y_1^n)- H(Y_2^n)+nc_7'\\
&\overset{(a)}{\leq} MnC + (M-1)H(Y_1^n) + n\cdot o(\log P)+nc_7',
\end{align*}
where $(a)$ is because  the difference $H(Y_1^n) - H(Y_2^n)$ can not be larger than $n\cdot o(\log{P})$ for the case with no eavesdropper's CSI from Section 6 of \cite{DavoodiJafar:14}.\footnote{The channel assumption in \cite{DavoodiJafar:14} is satisfied under our channel model.}
In terms of d.o.f., equivalently, we have 
\begin{align*}
d_s \leq \frac{M\alpha + M - 1}{M + 1}.
\end{align*}
Combining with the bound \eqref{eqn:cutset_multi} from the cutset bound, we finish the proof. \endproof

%%%%%%%%%%%%%%%%%%%%%%%%%%%%%%%%%%%%%%%%%%%%%%%%%%%

\subsection{Achievability} 
\subsubsection{Proof for the achievability part of Theorem \ref{thm2}}
Note that it is sufficient to show that the following two corner points are achievable:
$(\alpha, d_s)= \left( \frac{M  - 1}{M(M - 1) + 1}, \frac{M(M  - 1)}{M(M - 1) + 1}\right)$ and $(\alpha, d_s)=\left(\frac{1}{M} + \frac{2}{M^2}, 1\right)$. For the first corner point, the message with d.o.f. $\frac{M(M  - 1)}{M(M - 1) + 1}$ is split into $M$ independent partial messages each with d.o.f $\frac{M  - 1}{M(M - 1) + 1}$. The source sends each partial message to each different relay, which requires $\alpha=\frac{M  - 1}{M(M - 1) + 1}$. Then, the relays operate according to the cooperative jamming scheme in \cite{XieUlukus:14} for the Gaussian multiple access-wiretap channel. 

To show $(\alpha, d_s)=\left(\frac{1}{M} + \frac{2}{M^2}, 1\right)$ is achievable, we propose $\frac{M(M-1)}{2}$ sub-schemes, where the $(i,j)$-th sub-scheme for $i\in [1:M]$ and $j\in [1:M]$ such that $j>i$ achieves $\alpha_i=\alpha_j=\frac{2}{M}$, $\alpha_k=\frac{1}{M}$ for $k\notin \{i, j\}$, and $d_s=1$.  By time-sharing among these sub-schemes uniformly, we can prove that $(\alpha, d_s)=\left(\frac{1}{M} + \frac{2}{M^2}, 1\right)$ is achievable. Each sub-scheme is generalized from the S-AB scheme proposed in Section \ref{sec:achievability}. In Fig. \ref{fig:general}-(b), the (1,2)-th subscheme is illustrated for $M=4$. The message with d.o.f. $1$ is split into $M$ independent partial messages each with d.o.f $\frac{1}{M}$. In the $(i,j)$-th sub-scheme, the source sends each partial message to each different relay and sends a common noise with d.o.f. $\frac{1}{M}$ to relays $i$ and $j$ in addition to the partial messages, which requires $\alpha_i=\alpha_j=\frac{2}{M}$ and $\alpha_k=\frac{1}{M}$ for $k\notin \{i, j\}$. Then, each relay transmits what it has received in a way that the common noise signals are beam-formed in the null space of the legitimate destination's channel and the partial message signals are aligned with and are perfectly masked by the common noise signal at the eavesdropper. \endproof

\subsubsection{Proof for the achievability part of Theorem \ref{thm2_no}} 
Note that it is sufficient to show that the following two corner points are achievable: $(\alpha, d_s)=(\frac{M-1}{M^2}, \frac{M-1}{M})$ and $(\alpha, d_s)=(\frac{2}{M}, 1)$. First, $(\alpha, d_s)=(\frac{M-1}{M^2}, \frac{M-1}{M})$ can be shown to be achievable by  uniformly time-sharing $M$ sub-schemes, where the $k$-th sub-scheme for $k\in [1:M]$ achieves $\alpha_k=\frac{M-1}{M}$, $\alpha_j=0$ for $j\neq k$, and $d_s=\frac{M-1}{M}$. Each sub-scheme is a direct extension of the blind cooperative jamming scheme \cite{MukherjeeXieUlukus:arxiv15} for the wiretap channel with helpers, i.e., for the $k$-th sub-scheme, the source sends the message with d.o.f. $\frac{M-1}{M}$ to relay $k$ and the relays operate according to the blind cooperative jamming scheme \cite{MukherjeeXieUlukus:arxiv15} as if relay $k$ is the source and the other relays are the helpers. 

Next, $(\alpha, d_s)=(\frac{2}{M}, 1)$ can be shown to be achievable by uniformly time-sharing $\frac{M(M-1)}{2}$ sub-schemes, where the $(i,j)$-th sub-scheme for $i\in [1:M]$ and $j\in [1:M]$ such that $j>i$ achieves $\alpha_i=\alpha_j=1$, $\alpha_k=0$ for $k\notin \{i, j\}$, and $d_s=1$. Each sub-scheme is the same as the CoJ scheme proposed in Section \ref{sec:achievability}, i.e., in the $(i,j)$-th scheme, we use the CoJ scheme as if there are only two relays $i$ and $j$. \endproof

\begin{figure}[t]
 \centering
   \subfigure[]
  {\includegraphics[width=128mm]{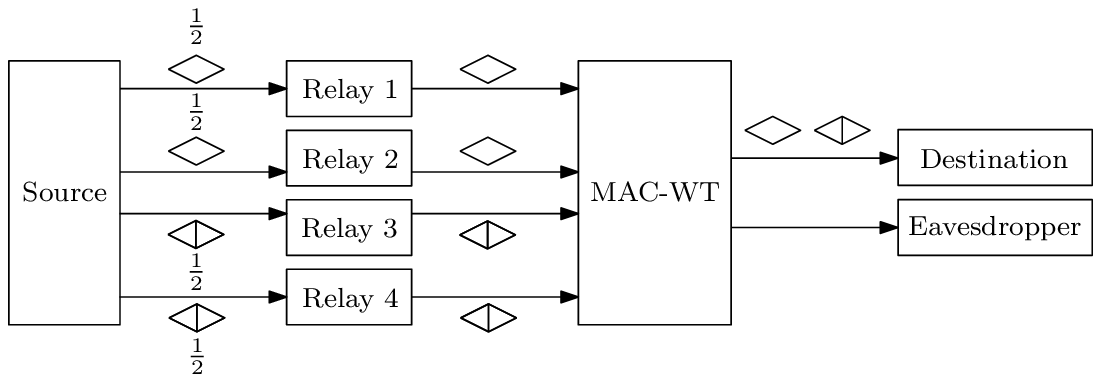}}
     \subfigure[]
  {\includegraphics[width=138mm]{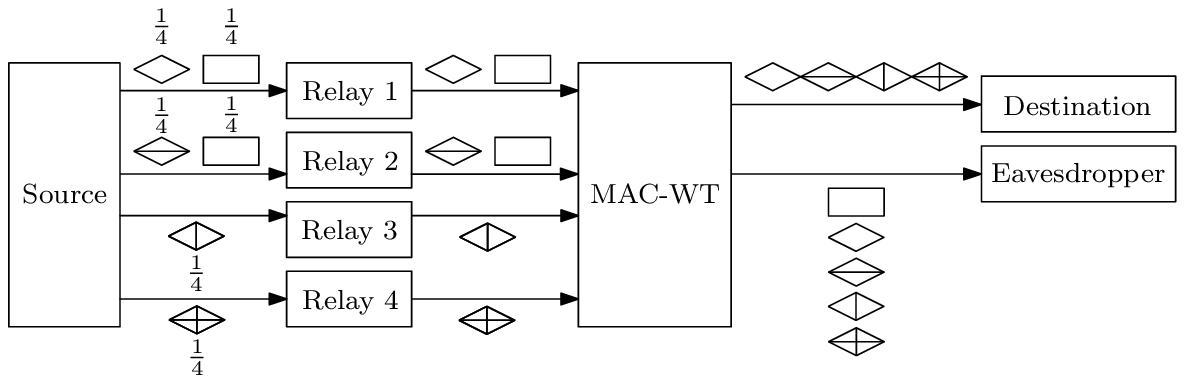}}
   \caption{Comparison between (a) the message-beamforming scheme and (b) the S-AB scheme for $M=4$. Similarly as in Fig. \ref{fig:with} and Fig. \ref{fig:no_csi}, diamond shapes and rectangular shapes represent (partial) messages and noises, respectively, with the number above or below each shape corresponding to its d.o.f. Same shapes with same patterns represent the same information, and otherwise independent informations. } \label{fig:general}
  \end{figure}

\begin{rem}
We note that a generalization of the message-beamforming scheme in Section \ref{sec:achievability} for the case with $M$-relays achieves $(\alpha, d_s)=(\frac{2}{M}, 1)$. Hence, the S-AB scheme outperforms the message-beamforming scheme for $M>2$. To see the intuition behind this, we illustrate some instances of using these two schemes for $M=4$ in Fig. \ref{fig:general}, where both the schemes achieve one secure d.o.f. but the S-AB scheme uses less link d.o.f.'s at the broadcast part.  For the message-beamforming scheme, every pair of two relays has to send a common partial message to beam-form each partial message. For the S-AB scheme, once two relays have common noise and independent partial messages as in the two-relay case, the other relays can send independent partial  messages without common noise since the same common noise can be used to mask all the partial messages simultaneously. Hence, the S-AB scheme requires less `common' information and thus is more efficient in the use of the broadcast links. 
\end{rem}

\section{Conclusion} \label{sec:conclusion}
In this paper, we established the exact secure d.o.f. of the Gaussian diamond-wiretap channel and generalized the results for multiple relays. We considered both the case with full CSI and the case with no eavesdropper's CSI, at the relays and the legitimate destination. Our results show that the absence of the eavesdropper's CSI reduces the secure d.o.f. for some range of moderate link d.o.f.'s of the broadcast part, but its effect decreases as the number of relays increases. 
For the converse part, we introduced a new technique of capturing the trade-off between the message rate and the amount of individual randomness injected at each relay. For the achievability part, we newly proposed a simultaneous alignment and beamforming (S-AB) scheme and a computation for jamming (CoJ) scheme  for the case with full CSI and for the case with no eavesdropper's CSI, respectively. Both the schemes incorporate transmitting common noise from the source to the relays and beamforming of common noise signals in the null space of the legitimate destination's channel. The S-AB scheme involves aligning the message and the common noise signals at the eavesdropper simultaneously with  the beamforming of the common noise signals. By doing so, it utilizes common information more efficiently than the message-beamforming scheme for more than two relays. The CoJ scheme involves computation between the message and the common noise symbols at the source, which requires less link d.o.f.'s at the broadcast part than  naively sending the message and the common noise separately. 

We note that our proposed schemes utilize the common information sent from the source to the two relays. It might be interesting to extend these schemes for the scenario where  the relays are allowed to conference to generate common noise or share common message, which is similar to the setting of \cite{ZhaoDingKhisti:15} for the diamond channel with no secrecy constraint. 
As a final remark, we note that our CoJ scheme can be useful in keeping the message secret from the relays.
Exploiting such a feature can be an interesting further work for the scenario where the source has common and confidential messages to each of the relays and the legitimate destination.

%\bibliography{../../../Reference/References}
%\bibliographystyle{IEEEtrans}

\end{document}